\documentclass[11pt,a4paper]{article}
\date{}
\usepackage{amsmath,amsthm}
\usepackage{graphicx}

\newtheorem*{t1}{Theorem 1}
\newtheorem*{t2}{Theorem 2}
\newtheorem*{t3}{Theorem 3}
\newtheorem*{t4}{Theorem 4}
\newtheorem*{t5}{Theorem 5}
\newtheorem*{t6}{Theorem 6}
\newtheorem*{t7}{Theorem 7}
\newtheorem*{t8}{Theorem 8}
\newtheorem*{t9}{Theorem 9}
\newtheorem*{t10}{Theorem 10}

\newtheorem*{l1}{Lemma 1}
\newtheorem*{l2}{Lemma 2}
\newtheorem*{l3}{Lemma 3}
\newtheorem*{l4}{Lemma 4}
\newtheorem*{l5}{Lemma 5}

\begin{document}

\markboth{Ch. Dangalchev}
{Closeness of some graph operations}

\title{Closeness of Some Graph Operations}

\author{Chavdar Dangalchev \\ dangalchev@hotmail.com
   \\ Institute of Mathematics and Informatics
   \\  Bulgarian Academy of Sciences, Sofia, Bulgaria. }

\maketitle

\begin{abstract}
Closeness is an important measure of network centrality.
In this article we will calculate the closeness of graphs, 
created by using operations on graphs.  
We will prove a formula for the closeness of shadow graphs.
We will calculate the closeness of line graphs of some 
well-known graphs (like path, star, cycle, and complete graphs) 
and the closeness of line graphs of two of these graphs, connected 
by a bridge (like lollipop, tadpole, broom, and bistar graphs).


\end{abstract}

\section{Introduction}

In network analysis, especially in social network analysis, finding the nodes which play a significant role is a very important task. How to find the most influential members in a social network, how to find the key nodes in the internet,  in utility networks, or in the spread of a disease, are main questions in many studies. In graph theory, most used central measures are degree, closeness, betweenness, and eigenvector. 

In this article we will focus on closeness centrality.
Dangalchev in [1] (research on network vulnerability) proposed the following definition for the closeness of vertex $i$ in simple undirected graphs:
\begin{equation}
C(i)=\sum\limits_{j\ne i} 2^{-d(i,j)}. 
\end{equation}

In the above formula,  $d(i,j)$  is the standard distance between vertices $i$  and $j$. The graph  $G$  closeness is the sum of all the vertices' closenesses:
\begin{equation}
\label{eq1}
C(G)=\sum\limits_i  \sum\limits_{j\ne i} 2^{-d(i,j)}. 
\end{equation}

The advantages of the above definition are that it can be used 
 for non-connected graphs, and it is convenient for creating formulae for graph operations
(see formulae (3), (4), and (5)).

Let vertex $p$ of graph $G_1$ (with vertex closeness $C(p)$ within graph $G_1$)  and vertex 
$q$  (with closeness $C(q)$)  of graph $G_2$ be connected by  link (edge) $(p,q)$ to create  graph $A$.
The formula for closeness of graph $A$ is given in [1]:
\begin{equation}
\label{eq3}
C(A )=C(G_{1} )+C(G_{2} )+\left( {1+C(p)} \right)\left( {1+C(q)} \right).
\end{equation}

Let vertices $p$  (with vertex closeness $C(p)$) of graph $G_1$ and  
$q$  (with closeness $C(q)$)  of graph $G_2$ coincide into one vertex 
$r$ to create graph $B$.
The formula for closeness of graph $B$  is given in [2]:
\begin{equation}
\label{eq4}
C(B)=C(G_{1} )+C(G_{2} )+2C(p)C(q).
\end{equation}
 Formulae (3) and (4) give us the opportunity to calculate the closeness of more complex networks by using the closeness of their parts.

 \quad 

\textbf{Definition 1: Shadow graph} 

\noindent Let graph $G$ have $n$  vertices $V = \{v_i:  i =  1,.., n\}$ and  edges  $(v_i, v_j) \in E(G)$. 
Shadow graph $S(G)$  
of graph $G$ is created by taking two copies $G_1$ and $G_2$ of $G$ with vertices 
$V_1 = \{v'_i:  i =  1,.., n\}$ and  $V_2 = \{v_i":  i =  1,.., n\}$. For every edge 
$(v_i, v_j)$  of $G$ we add 4 edges in $S(G)$: 
$(v'_i, v'_j)$, $(v'_i, v_j")$, $(v_i", v'_j)$, and $(v_i", v_j")$.

In section 2 we will prove the formula for the closeness of shadow graphs. 

 \quad 

\textbf{Definition 2: Line graph} 

\noindent Line graph $L(G)$ (another term used is edge-to-vertex dual graph) of an undirected graph $G$  is constructed in the following way: for each link of $G$, we add a vertex in $L(G)$; for every two links of $G$ that have a common vertex, we add an edge between their corresponding vertices in $L(G)$. 

In section 3 are given the closenesses of line graphs of some well-known graphs (like path, star, cycle, and complete graphs). In section 4 the closenesses of some line graphs with a bridge are calculated. 
In sections 5, 6, 7, and 8 are proven formulae for the closeness of lollipop, tadpole, broom, and bistar graphs, and for the closeness of their line graphs.

Closeness is a measure for centrality of a node. Another centrality measure is residual closeness - it is one of the most sensitive measures of graph vulnerability. The minimal closeness, after removing one link of a graph, is called link residual closeness (see [1]). It indicates the links critical for the existence of utility  or transportation networks. The minimal closeness, after removing one vertex and its links, is called vertex residual closeness [1]. It shows the most important members of a social network. The maximal closeness, after adding one link to a graph, is called additional closeness (see [2]). The additional closeness shows the growth potential of a network and indicates where to build new connections (in utility  or transportation networks).

Some results, related to closeness, additional closeness, and residual closeness can be found in [3-25]. Some of the articles focus on calculation of closeness or residual closeness of certain types of graphs: wheels in [3], Mycielski graphs in [6], caterpillar graphs in [7], splitting graphs in [9] and [14], thorn graphs in [10], banana trees in [17], Harary graphs in [24]. Other articles ([12], [15], [19],[22], [23]) show the extremal graphs (graphs with minimal or maximal closeness or residual closeness) when some characteristics of graphs are fixed.

\section{Calculating the closeness of shadow graphs}

In the next section we will calculate the closeness of shadow graphs.

The shadow graph $S(K_n)$ of complete graph $K_n$ with $n$ vertices
is complete graph $K_{2n}$ with deleted $n$ links $(v'_{i},v"_{j})$ between the 
corresponding vertices. All distances are equal to $1$, except for $2n$, which are 
equal to {2}.
\[
C(S(K_n)) = \frac{2n(2n-1)-2n}{2} + \frac{2n}{4}
=4 \cdot \frac{n(n-1)}{2} + \frac{n}{2}
=4C(K_n) + \frac{n}{2}.
\]

The shadow graph $S(S_n)$ of star graph $S_n$ with $n$ vertices
has $2$ center vertices at distance of $2$ to each other
and at distance of $1$ to $2n-2$ other vertices. It  also has $2n-2$ vertices,
corresponding to the star graph leaves,
at distance of $1$ to the two center vertices and
at distance of $2$ to the other $2n-3$ leaf vertices.
\begin{align}
\begin{split}\label{}
C(S(S_n))   &   =
2 \left( \frac{2n-2}{2} + \frac{1}{4} \right)
+ (2n-2) \left( \frac{2}{2} + \frac{2n-3}{4} \right)
\nonumber 
 \\  &  =
 \frac{2(4n-3)}{4} +   \frac{(2n-2)(2n+1)}{4} 
= \frac{8n-6 + 4n^2 -2n-2}{4}
 \\  &  =  4 \cdot \frac{n + n^2 -2}{4} + \frac{2n}{4} 
=  4 \cdot \frac{(n-1)( n+2)}{4} + \frac{n}{2}= 4C(S_n) + \frac{n}{2}.
\end{split}
\end{align}
Now let us consider path graphs.

\setlength{\unitlength}{.45in}
\begin{picture}(11,4.3)(-0.0,-1.5)

\put(3,0){\circle{0.08}}
\put(3,1){\circle{0.08}}
\put(4,0){\circle{0.08}}
\put(4,1){\circle{0.08}}
\put(5,0){\circle{0.08}}
\put(5,1){\circle{0.08}}
\put(6,0){\circle{0.08}}
\put(6,1){\circle{0.08}}
\put(7,0){\circle{0.08}}
\put(7,1){\circle{0.08}}

\linethickness{1pt}
\put(3,0){\line(1,1){1.0}}
\put(4,0){\line(1,1){1.0}}
\put(5,0){\line(1,1){1.0}}
\put(6,0){\line(1,1){1.0}}

\put(3,1){\line(1,-1){1.0}}
\put(4,1){\line(1,-1){1.0}}
\put(5,1){\line(1,-1){1.0}}
\put(6,1){\line(1,-1){1.0}}

\linethickness{0.7pt}

\put(3,0){\line(1,0){4.0}}
\put(3,1){\line(1,0){4.0}}

\put(3,-0.3){\makebox(0,0){1'}}
\put(4,-0.3){\makebox(0,0){2'}}
\put(5,-0.3){\makebox(0,0){3'}}
\put(6,-0.3){\makebox(0,0){4'}}
\put(7,-0.3){\makebox(0,0){5'}}

\put(3,1.3){\makebox(0,0){1"}}
\put(4,1.3){\makebox(0,0){2"}}
\put(5,1.3){\makebox(0,0){3"}}
\put(6,1.3){\makebox(0,0){4"}}
\put(7,1.3){\makebox(0,0){5"}}

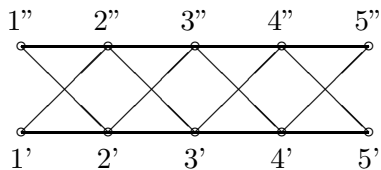
\put(5,-1.1){\makebox(0,0){Fig. 1: Shadow Graph $S(P_5)$.}}
\end{picture}

 In Fig. 1 is shown the shadow graph $S(P_5)$ of path graph $P_5$ with vertices  
$\{1,2,3,4,5\}$. We can see that every vertex is included into 2 path
graphs: for example, vertex $2'$ is part of paths $\{1',2',3',4',5'\}$ and
$\{1",2',3",4",5"\}$. The distance between vertices $2'$ and $2"$ is 2. 
Hence the closeness of the shadow graph $S(P_5)$ is:
\[
C(S(P_5))= 4 C(P_5) + 2 \cdot \frac {5}{4}.
\]
We can prove now:

\begin{t1}
 If $G$  is a connected graph with $n$ vertices, then the closeness of its shadow graph $S(G)$ is:
\begin{equation}
\label{eq5}
C (S (G) ) =  4C(G) + \frac {n}{2}.
\end{equation}
\end{t1}
\begin{proof}
We can follow the logic of the closeness of path graphs calculation.
\begin{align}
\begin{split}\label{}
 C ( S ( G) )  &   =
\sum\limits_{i=1}^n {\sum\limits_{j \ne i} {2^{-d(v'_{i},v'_{j})}} } +
\sum\limits_{i=1}^n {\sum\limits_{j=1}^n {2^{-d(v'_{i},v"_{j})}} } +
\nonumber 
 \\  &  \quad   + \sum\limits_{i=1}^n {\sum\limits_{i=1}^n {2^{-d(v"_{i},v'_{j})}} } +
\sum\limits_{i=1}^n {\sum\limits_{j \ne i} {2^{-d(v_{i}",v_{j}")}} }
 \\  & =  C(G) + \sum\limits_{i=1}^n {\left (2^{-d(v'_{i},v_{i}")} + C(v'_i) \right )} 
 \\  &  \quad   +
\sum\limits_{i=1}^n {\left (2^{-d(v_{i}",v_{i}')} + C(v_i") \right )}  + C(G)
 \\  & = C(G) + \frac{n}{4} + C(G) + \frac{n}{4} + C(G) + C(G) = 4 C(G) + \frac{n}{2},
\end{split}
\end{align}
which proves the Theorem.
\end{proof}

\section{Closeness of some well-known line graphs}

In this section we will calculate the closeness of line graphs of some well-known graphs.

The line graph of cycle graph $C_n$  with $n$ vertices is again cycle graph $C_n$:
\[
C(L(C_n)) =C(C_n).
\]
The line graph of path graph $P_n$  with $n$ vertices is path graph $P_{n-1}$ (the formula is proven in [1]):
\[
C(L(P_n)) =C(P_{n-1}) = 2n - 6 + 2^{3-n}. 
\]
The line graph of star $S_n$ graph with $n$ vertices is complete graph $K_{n-1}$  (the formula is proven in [1]):
\[
C(L(S_n)) =C(K_{n-1}) = \frac {(n-1)(n-2)}{2}. 
\]

For the closeness of the line graph of complete graph $K_n$  with $n$ vertices we will prove:
\begin{t2}
The closeness of line graph of complete graph $K_n$ is:
\begin{equation}
\label{eq6}
C(L(K_n)) = \frac{n (n ^3 + 2n^2 -13 n  + 10)}{16}
\end{equation}
\end{t2}
\begin{proof}Complete graph $K_n$ has $n(n-1) / 2$ links, so the number of vertices of the line graph $L(K_n)$ is $n(n-1) / 2$.

There are $n-1$ links at every vertex of the complete graph $K_n$,  which makes
$(n-1)(n-2) / 2$ pairs of these links. This means that every vertex of the complete graph $K_n$  creates 
$(n-1)(n-2) / 2$ links in the line graph $L(K_n)$, hence $n(n-1)(n-2) / 2$ is the total number of links in the line graph $L(K_n)$.

At every vertex of the complete graph, there are $n-1$ links. Every vertex of the line graph $L(K_n)$ is connected directly (a distance of 1)
to  $2(n-2)$ vertices, corresponding to both vertices of the link of the complete graph.

Every two links of the complete graph $K_n$, without common vertex, have 4 different links connecting their vertices.
Hence, the distance to the other  $n(n-1) / 2 - 1 - 2(n-2)$ vertices is 2. 

We can calculate the closeness of each vertex $i$ of line graph $L(K_n)$:
\begin{align}
\begin{split}
C(i) &  = \frac{2(n-2)}{2}  + \frac{1}{4} \left( n(n-1) / 2 - 1 - 2(n-2) \right )
\nonumber 
\\ &  = n-2 + \frac{1}{8} \left( n^2 -n  - 2  - 4n+8) \right )
\\ &  = \frac{8n - 16 + n^2 -5n  + 6}{8}  = \frac{ n ^2 + 3n -10}{8}
\end{split}
\end{align}
The closeness of the line graph is $n(n-1) / 2$ times bigger than  the closeness of vertex $i$:
\begin{align}
\begin{split}
C(L(K_n))  &  = \frac{ n ^2 + 3n -10}{8} \cdot \frac{n(n-1)}{2}
\nonumber 
\\ &  =  \frac{n}{16} \left( n ^3 + 3n^2 -10 n - n ^2 - 3n + 10 \right )
\\ &  =  \frac{n}{16} \left( n ^3 + 2n^2 -13 n  + 10 \right ),
\end{split}
\end{align}
and this finishes the proof.
\end{proof}

\section{Closeness of some line graphs with a bridge}

In the next sections we will calculate the closeness of line graphs of graphs, created by linking two graphs ($G_1$ and $G_2$) with bridge $B$. For the closeness we will follow formula (4):
\begin{equation}
\label{eq7}
C(L(G_1+B+G_2)) = C(L(G_1+B_1)) +C(L(G_2+B_2)) + 2C(B_1)C(B_2),
\end{equation}
where $L(G_1+B_1)$ is the line graph of graph $G_1$ and bridge $B_1$, and 
 $L(G_2+B_2)$ is the line graph of graph $G_2$ and bridge $B_2$. 
$C(B_1)$ is the closeness of vertex $B_1$ within line graph $L(G_1+B_1)$ and
$C(B_2)$ is the closeness of vertex $B_2$ within line graph $L(G_2+B_2)$. Next vertices $B_1$
and $B_2$ collapse into one bridge vertex $B$ to create line graph $L(G_1+B+G_2)$.

In this section we will prepare for the next sections by calculating the closeness of some line graphs of well-known graphs, 
together with a bridge link (see fig. 2). We will also calculate the closeness of the bridge vertex $C(B)$ within the line graph.

For path graph $P_n$ we can prove:
\begin{l1}
The closeness of line graph of  path $P_n$ and bridge link $B$  is:
\[
C(L(P_n + B)) = 2n - 4 + 2^{2-n}. 
\]
\end{l1}
\begin{proof}
If the bridge link $B$ is connected to a leaf vertex of the path $P_n$, then
the line graph of $P_n + B$  is  path $P_n$.
The closeness of $B$ within $L(P_n +B)$ is:
\[
C(B) = 2^{-1} + 2^{-2} + ... + 2^{1-n} = 1 -  2^{1-n}.
\]
The closeness of $P_n$ (see [1]) is:
\[
C(L(P_n + B)) = C(P_n) = 2n - 4 + 2^{2-n},
\]
which proves the Lemma.
\end{proof}

For cycle graph $C_n$ we have:
\begin{l2}
The closeness of line graph of  cycle $C_n$ and bridge link $B$  is:
\[
C(L(C_{2k}+B)) = 4k + 4 - (6k+4)2^ {-k},
\]
\[
C(L(C_{2k+1}+B)) = 4k + 6 - (8k+10)2^ {-k-1}.
\]
\end{l2}
\begin{proof}
The formulae  for closeness of cycle graphs, given in [1] and [3] are:
\begin{equation}
\label{eq8}
C(C_{2k}) = 4k - 6k2^ {-k},
\end{equation}
\begin{equation}
\label{eq9}
C(C_{2k+1}) = 2(2k+1) - 2(2k+1)2^ {-k}.
\end{equation}

The line graph of  cycle $C_n$ is again  cycle $C_n$. The bridge vertex $B$ 
is linked to two vertices and
\[
C(L(C_n+B)) = C(C_n) + 2 C(B).
\]
When $n=2k$ vertex $B$ has closeness:
\[
C(B) = 2(2^ {-1} + 2^ {-2}+ ...+ 2^ {-k}) = 2(1-2^ {-k}) = 2- 2^ {1-k}.
\]
\[
C(L(C_{2k} + B)) = 4k - 6k2^ {-k} + 2(2- 2^ {1-k}) = 4k + 4- (6k+4)2^ {-k}.
\]
When $n=2k+1$ vertex $B$ has closeness:
\[
C(B) = 2(2^ {-1} +  ...+ 2^ {-k}) + 2^ {-k-1} = 2- 2^ {1-k}+ 2^ {-k-1}=2- 3 \cdot 2^ {-k-1}.
\]
\[
C(L(C_{2k+1} + B)) = 2n- 2n2^ {-k}  + 2(2- 3 \cdot 2^ {-k-1}) = 4k + 6 - (8k+10)2^ {-k-1},
\]
which proves the formulae.
\end{proof}

For  star graph $S_n$ we have:
\begin{l3}
The closeness of line graph of  star $S_n$ and bridge link $B$, connected to a leaf  is:
\[
C(L(S_n + B)) =  \frac{(n-1)(n-2)+n}{2}.
\]
\end{l3}
\begin{proof}
The line graph of  star $S_n$ is complete $K_{n-1}$. Let the bridge vertex $B$ 
be linked to vertex $i$, not the center. Using formula (3), the closeness is:
\[
C(L(S_n + B)) = C(K_{n-1}) + (1+C(i))= \frac{(n-1)(n-2)}{2}+  \left(1+ \frac{(n-2)}{2}\right).
\]
The closeness of $B$ within $L(S_n +B)$ is:
\[
C(B) = 2^{-1} + (n-2) 2^{-2} = n 2^{-2} ,
\]
which proves the formula.
\end{proof}

We can prove:
\begin{l4}
The closeness of line graph of  star $S_n$ and bridge link $B$, connected to the center  is:
\[
C(L(S_n + B)) =  \frac{n(n-1)}{2}.
\]
\end{l4}
\begin{proof}
The line graph of  star $S_n$ is again complete $K_{n-1}$. The bridge vertex $B$ 
is linked to all other vertices.
The closeness of $B$ within $L(S_n +B)$ is:
\[
C(B) = (n-1) 2^{-1}.
\]
The closeness of $L(S_n +B)$ is:
\[
C(L(S_n + B)) = C(K_{n}) = \frac{n(n-1)}{2},
\]
which proves the formula.
\end{proof}

We will calculate the closeness of the line graph of the complete graph, 
together with a bridge link.

\setlength{\unitlength}{.45in}
\begin{picture}(11,4.3)(-0.0,-1.5)

\put(6.5,0){\circle{0.08}}
\put(6.5,1){\circle{0.08}}
\put(7.5,0){\circle{0.08}}
\put(7.5,1){\circle{0.08}}

\put(2.5,0){\circle{0.08}}
\put(2.5,1){\circle{0.08}}
\put(3.5,0){\circle{0.08}}
\put(4.5,0){\circle{0.08}}

\linethickness{1pt}
\put(6.5,0){\line(1,1){1.0}}
\put(2.5,1){\line(1,-1){1.0}}

\linethickness{0.7pt}
\put(6.5,0){\line(1,0){1.0}}
\put(6.5,1){\line(1,0){1.0}}
\put(6.5,0){\line(0,1){1.0}}
\put(7.5,0){\line(0,1){1.0}}

\put(2.5,0){\line(1,0){2.0}}
\put(2.5,0){\line(0,1){1.0}}

\put(6.5,-0.3){\makebox(0,0){3}}
\put(6.5,1.3){\makebox(0,0){2}}
\put(7.5,1.3){\makebox(0,0){1}}
\put(7.5,-0.3){\makebox(0,0){B}}

\put(3,-0.3){\makebox(0,0){3}}
\put(4,-0.3){\makebox(0,0){B}}
\put(2.2,0.5){\makebox(0,0){2}}
\put(3.4,0.5){\makebox(0,0){1}}

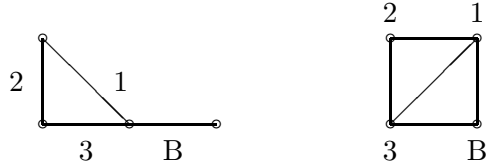
\put(5.0,-1.1){\makebox(0,0){Fig. 2: Graph $K_3$ with bridge B and its line graph.}}
\end{picture}

In Fig. 2, on the left  are complete graph $K_3$ and  link $B$. On the right is the line graph.
 The complete graph $K_3$ has links  $\{1,2,3\}$, which are the vertices of the line graph $L(K_3+B)$. Vertex $B$ corresponds to  bridge link $B$. 

\begin{l5}
The closeness of line graph of  complete graph $K_n$ and bridge link $B$ is:
\[
C(L(K_n+ B))  =   \frac{n ^4+ 2n^3 -9n^2 + 14n -8}{16}.
\]
\end{l5}
\begin{proof}
The line graph of complete graph $K_n$ and vertex $B$ has vertex $B$ connected to $n-1$ vertices of the line graph.

Using the proof of Theorem 2, vertex $B$ has $n-1$ vertices at distance $1$ and  $n(n-1) / 2 -n+1$ vertices at distance $2$. The closeness of vertex $B$ within graph $L(K_n)+B$ is:
\[
C(B)  = \frac{n-1}{2} + \frac{n(n-1) -2(n-1)}{8} = \frac{(n-1)(n+2)}{8}=  \frac{n ^2 + n  -2}{8}. 
\]
The closeness of $L(K_n+B)$ is:
\begin{align}
\begin{split}
C(L(K_n + B))&  = C(L(K_n)) + 2 \cdot \frac{n ^2 + n  -2}{8}
\nonumber 
\\ &  =  \frac{1}{16}\left(  n ^4+ 2n^3 -13 n^2  + 10n + 4 n^2 + 4 n -8 \right)
\\ &  =  \frac{n ^4+ 2n^3 -9n^2 + 14n -8}{16},
\end{split}
\end{align}
which proves the formula.
\end{proof} 

The closeness of the graph on the right in Fig. 2 is 5.5. The formula from Lemma 5 gives:
\[
C(L(K_3 + B)) =   \frac{3 ^4+ 2\cdot 3^3 -9 \cdot 3^2 + 14\cdot 3 -8}{16} = \frac{88}{16} =5.5.
\]

\section{Closeness of line graph of lollipop graphs}

The lollipop graph $L_{m,n}$  is created by connecting complete graph $K_m$ by a bridge to path graph $P_n$.

\setlength{\unitlength}{.45in}
\begin{picture}(11,4.3)(-0.0,-1.5)


\put(3.5,0){\circle{0.08}}
\put(3.5,2.0){\circle{0.08}}
\put(4.5,1){\circle{0.08}}
\put(5.5,1){\circle{0.08}}
\put(6.5,1){\circle{0.08}}

\linethickness{1pt}
\put(3.5,0){\line(1,1){1.0}}
\put(3.5,2.0){\line(1,-1){1.0}}
\put(3.5,0){\line(0,1){2.0}}

\linethickness{0.6pt}
\put(4.5,1){\line(1,0){2.0}}

\put(3.5,2.3){\makebox(0,0){4}}
\put(3.5,-0.3){\makebox(0,0){5}}

\put(4.5,1.3){\makebox(0,0){3}}
\put(5.5,1.3){\makebox(0,0){2}}
\put(6.5,1.3){\makebox(0,0){1}}

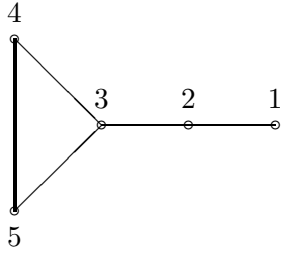
\put(5.2,-1.1){\makebox(0,0){Fig. 3: Lollipop Graph $L_{3,2}$.}}
\end{picture}

The complete graph $K_3$ in Fig. 3 has vertices  $\{3,4,5\}$,  the path graph $P_2$ has vertices  $\{1,2\}$, and the bridge is $(2,3)$.
We can calculate the closeness of lollipop graph $L_{m,n}$
by using formula (3):
\begin{t3}
The closeness of lollipop graph $L_{m,n}$ is:
\[
C(L_{m,n}) =  \frac {m}{2}  \left ( m+ 1 - 2^{1-n}  \right)
+ 2n-3 + 3\cdot 2^{-n}.
\]
\end{t3}
\begin{proof}
The closeness of complete graph $K_m$ is $m(m-1) / 2$, 
the closeness of path graph $P_n$ is $2n-4 + 2^{2-n}$.
The closeness of any vertex of $K_m$ is $(m-1) / 2$, 
the closeness of a leaf vertex of path graph $P_n$ is 
$1-2^{1-n}$. Applying formula (3) for lollipop graph $L_{m,n}$ we receive:

\begin{align}
\begin{split}
C(L_{m,n}) &  = \frac {m(m-1)}{2} + 2n-4 + 2^{2-n}
+ \left( 1 + \frac {m-1}{2} \right) 
\left( 1 + 1-2^{1-n} \right)
\nonumber 
\\ &  =  \frac {m(m-1)}{2} + 2n-4 + 2^{2-n} 
+  \frac {m+1}{2}
\left( 2-2^{1-n} \right)
\\ &  =  \frac {m(m+1)}{2} + 2n-3 + 2^{2-n} 
- m 2^{-n} -  2^{-n}
\\ &  =  \frac {m}{2}  \left ( m+ 1 - 2^{1-n}  \right)
+ 2n-3 + 3\cdot 2^{-n},
\end{split}
\end{align}
which finishes the proof.
\end{proof} 

The same formula is proven in [25].
We can prove now:
\begin{t4}
The closeness of the line graph of lollipop graph $L_{m,n}$ is:
\[
C(L(L_{m,n})) = \frac{m}{16}\left(  m ^3+ 2m^2 -5m   +18  \right)  -  (m ^2 +m -10) 2^{-n-1} + 2n - 5.
\]
\end{t4}
\begin{proof}
Using formula (7) and Lemmas 1 and 5,
the closeness of the line graph of lollipop graph $L_{m,n}$ is:

\begin{align}
\begin{split}
C(L(L_{m,n}))  &  = C(L(K_m + A))  + C(L(P_n+B))  + 2C(A)C(B)
\nonumber 
\\ &  =  \frac{m ^4+ 2m^3 -9m^2 + 14m -8}{16} + 2n - 4 + 2^{2-n} 
\\ &  \quad   + 2 \frac{m ^2 + m  -2}{8} (1 -  2^{1-n})
\\  &  =  \frac{1}{16}\left( m ^4+ 2m^3 -9m^2 + 14m -8 + 4m ^2 +4m -8  \right)
\\ &  \quad   + 2n - 4 + 2^{2-n} -  \frac{ m ^2 +m -2}{8} \cdot 2^{2-n}
\\  & =  \frac{1}{16}\left(  m ^4+ 2m^3 -5m^2   +18m - 16   \right)
\\ &  \quad    -  (m ^2 +m -10)  2^{-n-1} + 2n - 4,
\end{split}
\end{align}
which proves the theorem.
\end{proof}

\section{Closeness of tadpole graphs}

Tadpole graph $T_{m,n}$  is created by connecting cycle graph $C_m$ ($m \ge 3$)  by a bridge to path graph $P_n$.

\setlength{\unitlength}{.45in}
\begin{picture}(11,4.3)(-0.0,-1.5)


\put(3,1){\circle{0.08}}
\put(4,0){\circle{0.08}}
\put(4,2.0){\circle{0.08}}
\put(5,1){\circle{0.08}}
\put(6,1){\circle{0.08}}
\put(7,1){\circle{0.08}}

\linethickness{1pt}
\put(4,0){\line(1,1){1.0}}
\put(4,2.0){\line(1,-1){1.0}}
\put(4,0){\line(-1,1){1.0}}
\put(4,2.0){\line(-1,-1){1.0}}

\linethickness{0.6pt}
\put(5,1){\line(1,0){2.0}}

\put(4,2.3){\makebox(0,0){4}}
\put(4,-0.3){\makebox(0,0){6}}
\put(2.7,1.0){\makebox(0,0){5}}

\put(5,1.3){\makebox(0,0){3}}
\put(6,1.3){\makebox(0,0){2}}
\put(7,1.3){\makebox(0,0){1}}

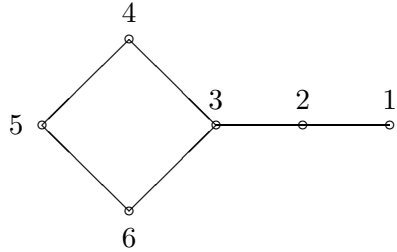
\put(5.2,-1.1){\makebox(0,0){Fig. 4: Tadpole Graph $T_{4,2}$.}}
\end{picture}

The cycle graph $C_4$ in Fig. 4 has vertices  $\{3,4,5,6 \}$, the path graph $P_2$ has vertices  $\{1,2\}$, and the bridge is $(2,3)$. 
We can calculate the closeness of tadpole graph $T_{m,n}$
by using formula (3):
\begin{t5}
The closeness of tadpole graph $T_{m,n}$ is:
\[
C(T_{2k,n}) = 4k -6k 2^ {-k} + 2n+2  
+ 6 \left( 2^{-n-k} -2^ {-k}\right) - 2^{1-n},
\]
\[
C(T_{2k+1,n}) =  4k -k2^{2-k}+ 2n + 4
+ 2^{2-n-k} - 3 \cdot 2^{1-k}  -2^{1-n}.
\]
\end{t5}
\begin{proof}
The closeness of the cycle graph $C_m$ is $C(C_m)$, 
the closeness of the path graph $P_n$ is $2n-4 + 2^{2-n}$.
The closeness of a vertex of $C_m$ is $C(C_m) / m$, 
the closeness of leaf vertex of path graph $P_n$ is 
$1-2^{1-n}$. Applying formula (3) for tadpole graph $T_{m,n}$ we receive:

\begin{align}
\begin{split}
C(T_{m,n}) &  = C(C_m) + 2n-4 + 2^{2-n}
+ \left( 1 + \frac {C(C_m)}{m} \right) 
\left( 1 + 1-2^{1-n} \right)
\nonumber 
\\ &  =  C(C_m) \frac {m+2-2^{1-n}}{m} + 2n-2 + 2^{1-n}.
\end{split}
\end{align}
 When $m=2k$, using formula (8), the tadpole graph closeness is:
\begin{align}
\begin{split}
C(T_{2k,n}) &  = \left( 4k - 6k2^ {-k} \right) \frac {2k+2-2^{1-n}}{2k} + 2n-2 + 2^{1-n}
\nonumber 
\\ &  = \left( 2 - 3 \cdot 2^ {-k} \right)  \left( 2k+2-2^{1-n}\right)  + 2n-2 + 2^{1-n}
\\ &  = 4k -2^{2-n} - 6(k+1) 2^ {-k} + 3 \cdot 2^{1-n-k}
+ 2n+2 + 2^{1-n}
\\ &  = 4k -6k 2^ {-k} + 2n+2 - 3 \cdot 2^ {1-k}  
+ 3 \cdot 2^{1-n-k} - 2^{1-n}.
\end{split}
\end{align}

 When $m=2k+1$, using formula (9), the tadpole graph closeness is:
\begin{align}
\begin{split}
C(T_{2k,n}) &  =  \left(2(2k+1) - 2(2k+1)2^ {-k} \right) \frac {2k+3-2^{1-n}}{2k+1} + 2n-2 + 2^{1-n}
\nonumber 
\\ &  = \left(2 - 2^ {1-k} \right) \left(2k+3-2^{1-n}\right) + 2n-2 + 2^{1-n}
\\ &  = 4k -2^{1-n}
 - (2k+3)2^{1-k} + 2^{2-n-k} + 2n + 4.
\end{split}
\end{align}
which finishes the proof.
\end{proof}

For the closeness of the line graph, we can prove now:
\begin{t6}
The closeness of line graph of tadpole graph $T_{m,n}$ is:
\[
C(L(T_{2k,n})) = 4k + 2n  - (6k+8)2^ {-k}  - 2^{2-n} + 4 + 2^{3-k-n},
\]
\[
C(L(T_{2k+1,n})) = 4k + 2n + 6 - (4k+8)2^{-k} - 2^{2-n} + 3 \cdot 2^{1-k-n}.
\]
\end{t6}
\begin{proof}
Using formula (7) and Lemmas 1 and 2:
\begin{align}
\begin{split}
C(L(T_{m,n}))   &  = C(L(C_m + A) ) + C(L(P_n+B))  +2C(A)C(B)
\nonumber 
\\ &  =  C(L(C_m + A))  + 2n - 4 + 2^{2-n}  +2(1 -  2^{1-n})C(A)
\end{split}
\end{align}
 When $m=2k$ the line graph of tadpole graph closeness:
\begin{align}
\begin{split}
C(L(T_{2k,n}))   &  = 4k + 4- (6k+4)2^ {-k}  + 2n - 4 
\nonumber 
\\  & \quad + 2^{2-n}  +(2 -  2^{2-n}) (2- 2^ {1-k})
\\ &  =  4k + 2n  - (6k+8)2^ {-k}  - 2^{2-n} + 4 + 2^{3-k-n}.
\end{split}
\end{align}

When $n=2k+1$ the closeness is:
\begin{align}
\begin{split}
C(L(T_{2k+1,n}))   &  = 4k + 6 - (8k+10)2^ {-k-1}  + 2n - 4 
\nonumber 
\\  & \quad + 2^{2-n}  +(2 -  2^{2-n})(2- 3 \cdot 2^ {-k-1})
\\ &  =  4k + 2n + 6 - (8k+16)2^{-k-1} - 2^{2-n} + 3 \cdot 2^{1-k-n},
\end{split}
\end{align}
which proves the formulae.
\end{proof} 

\section{Closeness of broom graphs}
Broom graph $B_{m,n}$  is created by connecting path graph $P_n$ by a bridge to the center of star graph $S_m$ ($m \ge 3$). The case, when the bridge is connected to a leaf of the star graph, is considered similarly, only $m$ and $n$ are different.

\setlength{\unitlength}{.45in}
\begin{picture}(11,4.3)(-0.0,-1.5)

\put(3.5,1){\circle{0.08}}
\put(4,0){\circle{0.08}}
\put(4,2.0){\circle{0.08}}
\put(5,1){\circle{0.08}}
\put(6,1){\circle{0.08}}
\put(7,1){\circle{0.08}}

\linethickness{1pt}
\put(4,0){\line(1,1){1.0}}
\put(4,2.0){\line(1,-1){1.0}}

\linethickness{0.6pt}
\put(3.5,1){\line(1,0){3.5}}

\put(4,2.3){\makebox(0,0){4}}
\put(4,-0.3){\makebox(0,0){6}}
\put(3.2,1.0){\makebox(0,0){5}}

\put(5,1.3){\makebox(0,0){3}}
\put(6,1.3){\makebox(0,0){2}}
\put(7,1.3){\makebox(0,0){1}}

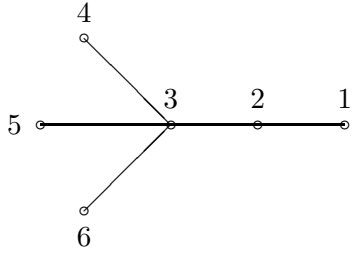
\put(5.2,-1.1){\makebox(0,0){Fig. 5: Broom Graph $B_{4,2}$.}}
\end{picture}

The star graph $S_4$ in Fig. 5 has vertices  $\{3,4,5,6 \}$, the path graph $P_2$ has vertices  $\{1,2\}$, and the bridge is $(2,3)$.
We can calculate the closeness of broom graph $B_{m,n}$
by using formula (3):
\begin{t7}
The closeness of broom graph $B_{m,n}$ is:
\[
C(B_{m,n}) = \frac {m}{4}  \left ( m+ 5 - 2^{2-n}  \right)
+ 2n-3.5 + 3\cdot 2^{-n}.
\]
\end{t7}
\begin{proof}
The closeness of star graph $S_m$  (see[1]) is $(m-1)(m+2) / 4$, 
and the closeness of path graph $P_n$ is $2n-4 + 2^{2-n}$.
The closeness of the center of $S_m$ is $(m-1) / 2$, 
the closeness of a leaf vertex of path graph $P_n$ is 
$1-2^{1-n}$. Applying formula (3) for broom graph $B_{m,n}$ we receive:
\begin{align}
\begin{split}
C(B_{m,n}) &  = \frac {(m-1)(m+2)}{4} + 2n-4 + 2^{2-n}
+ \left( 1 + \frac {m-1}{2} \right) 
\left( 1 + 1-2^{1-n} \right)
\nonumber 
\\ &  =  \frac {m(m+1)}{4} + 2n-4.5 + 2^{2-n} 
+  \frac {m+1}{2}
\left( 2-2^{1-n} \right)
\\ &  =  \frac {m(m+5)}{4} + 2n-3.5 + 2^{2-n} 
- m 2^{-n} -  2^{-n}
\\ &  =  \frac {m}{4}  \left ( m+ 5 - 2^{2-n}  \right)
+ 2n-3.5 + 3\cdot 2^{-n},
\end{split}
\end{align}
which finishes the proof.
\end{proof} 

For the line graph we have:
\begin{t8}
The closeness of the line graph of broom graph $B_{m,n}$ is:
\[
C(L(B_{m,n})) = m(m+1)2^{-1} + 2n -5 + (3-m)2^{1-n}.
\]
\end{t8}
\begin{proof}
Using formula (7) and Lemmas 1 and 4:
\begin{align}
\begin{split}
C(L(B_{m,n}))   &  = C(L(S_m + A))  + C(L(P_n+B))  +2C(A)C(B)
\nonumber 
\\ &  =  \frac{m(m-1)}{2}  + 2n - 4 + 2^{2-n}  +2\frac{m-1}{2}(1 -  2^{1-n})
\\   &  =  \frac{m^2 - m + 2m}{2} + 2n - 4 -1 + 2^{1-n} (2-m+1)
\\ &  =  \frac{m(m+1)}{2}  + 2n +  (3-m) 2^{1-n}-5,
\end{split}
\end{align}

which proves the formula.
\end{proof}

\section{Closeness of bistar graphs}
Bistar graph $BS_{m,n}$  is created by connecting the center of star graph $S_m$ ($m \ge 3$) by a bridge to 
 the center of star graph $S_n$ ($n \ge 3$).

\setlength{\unitlength}{.45in}
\begin{picture}(11,4.3)(-0.0,-1.5)

\put(3.5,1){\circle{0.08}}
\put(4,0){\circle{0.08}}
\put(4,2.0){\circle{0.08}}
\put(5,1){\circle{0.08}}
\put(6,1){\circle{0.08}}
\put(7,0){\circle{0.08}}
\put(7,2){\circle{0.08}}

\linethickness{1pt}
\put(4,0){\line(1,1){1.0}}
\put(4,2.0){\line(1,-1){1.0}}
\put(6,1){\line(1,1){1.0}}
\put(6,1){\line(1,-1){1.0}}

\linethickness{0.6pt}
\put(3.5,1){\line(1,0){2.5}}

\put(4,2.3){\makebox(0,0){4}}
\put(4,-0.3){\makebox(0,0){6}}
\put(3.2,1.0){\makebox(0,0){5}}

\put(5,1.3){\makebox(0,0){3}}
\put(6,1.3){\makebox(0,0){2}}
\put(7,2.3){\makebox(0,0){1}}
\put(7,-0.3){\makebox(0,0){7}}

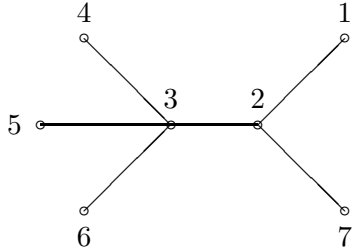
\put(5.2,-1.1){\makebox(0,0){Fig. 6: Bistar Graph $BS_{4,3}$.}}
\end{picture}

 The star graph $S_4$ in Fig. 6 has vertices  $\{3,4,5, 6\}$, the star graph $S_3$ has vertices  $\{1,2,7\}$, and the bridge is $(2,3)$.

We can calculate the closeness of bistar graphs using formula (3):
\begin{t9}
The closeness of bistar graph $BS_{m,n}$ is:
\[
C(BS_{m,n}) =  \frac {1}{4} \left( m(m+2) + n(n+2) +mn -3 \right).
\]
\end{t9}
\begin{proof}
The closeness of star graph $S_m$ (see [1]) is $(m-1)(m+2) / 4$, 
and the closeness of star graph $S_n$ is $(n-1)(n+2) / 4$.
The closeness of the center of $S_m$ is $(m-1) / 2$, 
and the closeness of the center of $S_n$ is $(n-1) / 2$. Applying formula (3) for bistar graph $BS_{m,n}$ we receive:

\begin{align}
\begin{split}
C(BS_{m,n}) &  = \frac {(m-1)(m+2)}{4} + \frac {(n-1)(n+2)}{4}
\nonumber 
\\   &  \quad  
+ \left( 1 + \frac {m-1}{2} \right) 
\left( 1 + \frac {n-1}{2} \right) 
\\ &  =  \frac {1}{4} \left( m(m+1) + n(n+1) - 4+(m+1)(n+1)  \right)
\\ &  =  \frac {1}{4} \left( m(m+2) + n(n+2) +mn-3  \right),
\end{split}
\end{align}
which finishes the proof.
\end{proof} 

We can prove for the line graph of bistar graphs:
\begin{t10}
The closeness of the line graph of bistar graph $BS_{m,n}$ is:
\[
C(L(BS_{m,n})) = \frac{1}{2} \left( (m-1)^2  + (n-1)^2 +mn  -1 \right).
\]
\end{t10}
\begin{proof}
Using formula (7) and Lemma 4:
\begin{align}
\begin{split}
C(L(BS_{m,n}))   &  = C(L(S_m + A))  + C(L(S_n+B))  +2C(A)C(B)
\nonumber 
\\ &  =  \frac{m(m-1)}{2}  +  \frac{n(n-1)}{2}  +2\frac{m-1}{2} \cdot \frac{n-1}{2}
\\   &  =  \frac{m^2 - m + n^2 - n +mn - m -n +1}{2} 
\\ &  =  \frac{(m-1)^2  + (n-1)^2 +mn  -1}{2},
\end{split}
\end{align}
which proves the formula.
\end{proof} 

\section{Conclusion} 

In this article, we have proven a formula for the closeness of shadow graphs. 
We have also calculated the closenesses 
 of line graphs of some well-known graphs, like path, star, cycle,
and complete graphs.
The closenesses of  graphs, created by linking two of these graphs by a bridge  (like lollipop, tadpole, broom, and bistar graphs) are given.
The closenesses of line graphs of these graphs are calculated.

Future work should focus on calculating closeness, residual closeness, and additional closeness of more complex graphs.

\end{document}